\documentclass[draft,grl]{agutex}

\usepackage{amsfonts,amsmath,amsbsy}
\usepackage{graphicx}

\setkeys{Gin}{draft=false}

\newcommand{\figref}[1]{Fig.\ \ref{#1}}

\def\pct{\%{ }}

\def\citeapos#1{\citeauthor{#1}'s [\citeyear{#1}]}
\makeatletter
\renewcommand*\env@matrix[1][\arraystretch]{%
  \edef\arraystretch{#1}%
  \hskip -\arraycolsep
  \let\@ifnextchar\new@ifnextchar
  \array{*\c@MaxMatrixCols c}}
\makeatother

\authorrunninghead{Wang}%
\titlerunninghead{Coherent water transport}%
\lefthead{Wang}%
\righthead{Coherent water transport}%
\journalid{2015}%
\articleid{X}{X}%
\paperid{2015JGRXXXXX}%
\cpright{AGU}{2015}%
\received{\today}%
\revised{\today}%
\accepted{\today}%

\authoraddr{Y.\ Wang and M.\ J. Olascoaga, RSMAS/OCE, University
of Miami, 4600 Rickenbacker Cswy., Miami, FL 33149, USA.
(ywang@rsmas.miami.edu; jolascoaga@rsmas.miami.edu)}

\authoraddr{F.\ J. Beron-Vera, RSMAS/ATM, University
of Miami, 4600 Rickenbacker Cswy., Miami, FL 33149, USA.
(fberon@rsmas.miami.edu)}

\begin{document}

\title{Coherent water transport across the South
Atlantic}

\bigskip \authors{Y.\ Wang\altaffilmark{1}, M.\ J.\
Olascoaga\altaffilmark{1}, F.\ J.\ Beron-Vera\altaffilmark{2}}

\altaffiltext{1}{Department of Ocean Sciences, Rosenstiel School
of Marine and Atmospheric Science, University of Miami, Miami,
Florida, USA.} \altaffiltext{2}{Department of Atmospheric Sciences,
Rosenstiel School of Marine and Atmospheric Science, University of
Miami, Miami, Florida, USA.}

\begin{abstract}
   The role of mesoscale eddies in transporting Agulhas leakage is
   investigated using a recent technique from nonlinear dynamical
   systems theory applied on geostrophic currents inferred from the
   over two-decade-long satellite altimetry record.  Eddies are
   found to acquire material coherence away from the Agulhas
   retroflection, near the Walvis Ridge in the South Atlantic.
   Yearly, 1 to 4 coherent material eddies are detected with diameters
   ranging from 40 to 280 km.  A total of 23 eddy cores of about
   50 km in diameter and with at least 30\pct of their contents
   traceable into the Indian Ocean were found to travel across the
   subtropical gyre with minor filamentation.  Only 1 eddy core was
   found to pour its contents on the North Brazil Current. While
   ability of eddies to carry Agulhas leakage northwestward across
   the South Atlantic is supported by our analysis, this is more
   restricted than suggested by earlier ring transport assessments.
\end{abstract}

\begin{article}

\section{Introduction}

Mesoscale eddies are widely recognized as potential agents of
long-range water transport [e.g., \emph{Robinson},
1983]\nocite{Robinson-83}.  Agulhas rings, in particular,
have long been thought as conduits for the leakage of warm and salty
Indian Ocean water into the South Atlantic \citep{DeRuijter-etal-99,
Gordon-86, Lutjeharms-06, Richardson-07, vanSebille-vanLeeuwen-07}
and as such contributors to the maintenance of the meridional
overturning circulation in the Atlantic \citep{Gordon-86, Weijer-etal-02,
Knorr-Lohmann-03, Peeters-etal-04, Beal-etal-11}.  The role of rings
in carrying Agulhas leakage was emphasized by \citet{Gordon-Haxby-90},
who argued that the rings, after being shed from the Agulhas
retroflection as the result of occasional Indian-Ocean-entrapping
occlusions, travel across the South Atlantic and pour their contents
on the North Brazil Current.  While this long-range transport view
on rings has been challenged by the potentially important role of
filaments and other forms of transport \citep{Schouten-etal-00,
vanSebille-etal-10a}, it remains to be tested using tools designed
to unambiguously frame the structures of interest: eddies with
persistent material cores.  Here we extract from geostrophic currents
inferred from the over two-decade-long record of satellite altimetry
measurements of sea surface height (SSH) mesoscale
coherent material eddies, investigate their
life cycles, construct a time series of coherent water transport,
and evaluate the significance of the obtained transport estimates.
The eddies are extracted using a recently developed Lagrangian
method from nonlinear dynamical systems theory \citep{Haller-Beron-13,
Haller-Beron-14}.  Independent of the reference frame chosen, the
method allows identification and tracking, in forward direction
until the time of their demise and in backward direction to the
time of their genesis, of eddies shielded by extraordinarily resilient
fluid belts that defy the exponential stretching of typical fluid
belts in turbulence.  Such material eddies coherently
transport the enclosed fluid with no noticeable leakage through the
flow domain for the whole extent of their lifetimes.  Eddies revealed
from their Eulerian footprints , as is the case of the rings
considered by \citet{Gordon-Haxby-90},  do not possess this property
\citep{Beron-etal-13}, which is critical to assess the validity of
\citeapos{Gordon-Haxby-90} long-range transport view on Agulhas
rings.

\section{Methods}

\citet{Haller-Beron-13, Haller-Beron-14} consider
exceptional material loops in turbulent flow that
form the centerpieces of thin material belts exhibiting no
leading order change in averaged stretching as the widths of the
belts are varied. Solutions to this variational problem are material
loops such that each of their subsets are stretched by a unique
factor $\lambda$ when the loops are advected from time $t_0$ to
time $t$.  Being uniformly stretching, these $\lambda$-loops resist
the exponential stretching typical material loops experience in
turbulence.  Represented as closed curves $s \mapsto x_0(s)$, where
parameter $s$ is periodic, the $\lambda$-loops satisfy one of the
two equations:
\begin{equation}
  \frac{\mathrm{d}x_0}{\mathrm{d}s} = 
  \sqrt{
  \frac
  {\lambda_2(x_0) - \lambda^2}
  {\lambda_2(x_0) - \lambda_1(x_0)}
  }
  \,\xi_1(x_0) 
  \pm
  \sqrt{
  \frac
  {\lambda^2 - \lambda_1(x_0)}
  {\lambda_2(x_0) - \lambda_1(x_0)}
  }
  \,\xi_2(x_0).  
  \label{eq:eta}
\end{equation}
Here  $0 < \lambda_1(x_0) \leq \lambda_2(x_0)$ and $\xi_i(x_0) \cdot
\xi_j(x_0) = \delta_{ij}$ are eigenvalues and (normalized) eigenvectors,
respectively, of the right Cauchy--Green strain tensor field,
$C_{t_0}^t(x_0) := \mathrm{D}F_{t_0}^t(x_0) ^\top
\mathrm{D}F_{t_0}^t(x_0)$, a frame-invariant (or objective) measure
of deformation where $F_{t_0}^t(x_0) := x(t;x_0,t_0)$ is the flow
map that associates times $t_0$ and $t$ positions of fluid particles,
which evolve according to
\begin{equation}
  \frac{\mathrm{d}x}{\mathrm{d}t} = v(x,t),
  \label{eq:dxdt}
\end{equation}
where $v(x,t)$ is a two-dimensional velocity field.  Closed curves
satisfying \eqref{eq:eta} occur in families of nonintersecting limit
cycles, necessarily encircling singularities of $C_{t_0}^t(x_0)$,
i.e., points where the field is isotropic.  The outermost member
of a family of $\lambda$-loops will be observed physically as the
boundary of a \emph{coherent material eddy}: immediately outside,
no coherent belt may exist containing the eddy.  Limit cycles of
\eqref{eq:eta} tend to exist only for $\lambda \approx 1$.  Material
loops characterized by $\lambda = 1$ reassume their initial arclength
at time $t$.  This property, along with conservation of enclosed
area in the incompressible case, creates extraordinary coherence.

Coherent material eddy detection and tracking is implemented as
follows (detailed algorithm steps are given in the Appendix of
\citet{Haller-Beron-13}).  1) Fix a domain $U$ and a time scale $T$
over which eddies are to be identified.  2) On $U$ set a grid
$\mathcal{G}_U$ of initial positions $x_0$.  3) For each $x_0 \in
\mathcal{G}_U$ integrate \eqref{eq:dxdt} from $t_0$ to $t = t_0 +
T$, obtaining a discrete approximation of $F_{t_0}^t(x_0)$.  4)
Evaluate $\mathrm{D}F_{t_0}^t(x_0)$ using finite differences, then
construct $C_{t_0}^t(x_0)$, and finally compute $\{\lambda_i(x_0)\}$
and $\{\xi_i(x_0)\}$.  5) Locate eddy candidate regions $V$ by
isolating singularities of $C_{t_0}^t(x_0)$ surrounded by
singularity-free annular regions.  6) In each $V$ repeat the first
two steps using a finer grid $\mathcal{G}_V$, and seek the outermost
possible limit cycle of \eqref{eq:eta} with the aid of a Poincare
section (a limit cycle corresponds to a fixed point of the Poincare
map) starting with $\lambda = 1$.  If no limit cycle is found for
any $\lambda$ (typically near 1), the candidate region does not
contain a coherent material eddy.  7) Finally, advect the boundary
of the coherent material eddy detected to track its motion.

Here we have specifically considered $v(x,t) =
gf^{-1}\nabla^\perp\eta(x,t)$, where $g$ is the acceleration of
gravity, $f$ stands for Coriolis parameter, $\perp$ represents a
90$^{\circ}$-anticlockwise rotation, and $\eta(x,t)$ is the SSH,
taken as the sum of a (steady) mean dynamic topography and the
(transient) altimetric SSH anomaly, both distributed by AVISO
(Archiving, Validation and Interpretation of Satellite Oceanographic
Data); specific products employed are Rio05 and DT-MSLA ``all sat
merged,'' respectively.  The mean dynamic topography is constructed
from satellite altimetry data, in-situ measurements, and a geoid
model \citep{Rio-Hernandez-04}.  The SSH anomaly is provided weekly
on a 0.25$^{\circ}$-resolution longitude--latitude grid.  This is
referenced to a 20-year (1993--2012) mean, obtained from the combined
processing of data collected by altimeters on the constellation of
available satellites \citep{LeTraon-etal-98}. Here the weekly SSH
fields are interpolated daily, which reduces trajectory overshooting
\citep{Keating-etal-11}. We chose $U = [20.5^{\circ}\text{W},
10.5^{\circ}\text{E}] \times [29.5^{\circ}\text{S},32.5^{\circ}\text{S}]$
(indicated by a box in \figref{fig:fig01}, bottom-left panel).  This
domain intersects the so-called Agulhas corridor \citep{Goni-etal-97}.
It lies sufficiently away from the Agulhas retroflection to allow
coherence to build up, and is sufficiently wide, both zonally, to
capture all eddies possibly shed, and meridionally, to fit the
largest such eddies.  We set $T = 90$, 180, and 360 d.  This resulted
in detections of eddies with maximum diameters decreasing from
around 280 km to 100 km.  Eddy diameters remained stable for $T$
in the range 30--90 d; for $T$ shorter than 30 d or longer than 360
d coherence was difficult to be revealed on any scale.  Detections
were carried out  over 1992--2013 (nearly the entire period of
available altimetry measurements) in such a way that $U$ was filled
with new eddies at each $t_0$, thereby avoiding defective or redundant
eddy counting.  We set $\mathcal{G}_U$ and $\mathcal{G}_V$ to be
regular with square elements of roughly 1.5- and 1-km side,
respectively.  All integrations were carried out using a stepsize-adapting
fourth-order Runge--Kutta method with interpolations obtained using
a cubic scheme.  In the case of \eqref{eq:eta}, further care had
to be taken by enforcing a unique eigenvector field orientation at
each integration step.

\section{Results}

\begin{figure}
  \includegraphics[width=\textwidth]{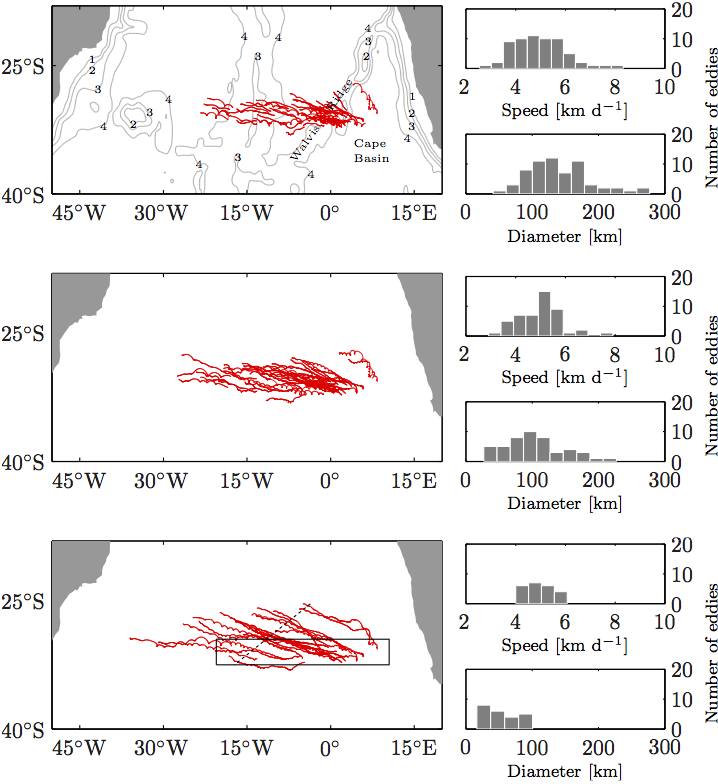}%
  \caption{Trajectories (left column) and mean translation speeds
  and diameters (right column) of coherent material eddies in the
  Agulhas corridor as detected from altimetry-derived velocities
  over 1992--2013 with lifetimes 90 (top row), 180 (middle row),
  and 360 (bottom row) d. Indicated in the bottom-left panel are
  detection domain (solid rectangle) and the reference section used
  in the construction of the coherent transport time series of
  \figref{fig:fig02} (dashed line). Selected bathymetry levels (in
  km) are indicated in the top-left panel along with two relevant
  topographic features.}
  \label{fig:fig01}%
\end{figure}

We begin by showing in \figref{fig:fig01} trajectories (left column)
and histograms of mean translational speeds and diameters (right
column) of coherent material eddies detected from $T = 90$ (top
row), 180 (middle row), and 360 (bottom row) d integrations.  A
total of 59 (4), 47 (1), and 23 (0) anticyclonic (cyclonic) eddies
are detected over 1992--2013.  [Ignored from the analysis only are
7 (6), 3 (3), and 1 (0) anticyclones (cyclones) that take southwestward
directions as these eddies are not relevant for our purposes here.]
The predominance of anticyclones over cyclones signals enhanced
stability for anticyclones in agreement with prior results
\citep{vanSebille-etal-10b}.  Of these eddies 39, 40, and 19 are
found with $\lambda = 1$; for all other eddies $\lambda$ ranges
from 0.9 to 1.1. It must be realized that 180- and 360-d eddies lie
inside 90-d eddies at detection time, i.e.,  180- and 360-d eddies
do not constitute different eddies but rather constitute 90-d eddy
coherent material cores.  In effect, an eddy boundary detected with
a given $T$ typically lies quite close to some member of the family
of $\lambda$-loops that fill an eddy detected with a shorter $T$.
The detection rate is quite irregular.  It varies from 1 to 4 eddies
per year.  This applies to 90- and 180-d eddies; it applies to 360-d
eddies too when they are present, namely, all years except 1994--1995,
2004, and 2006--2007.  The irregularity of the detection rate is
indicative of substantial coherent material eddy episodicity rather
than an artifact created by the altimetry set.  Indeed, while earlier
years are covered by fewer satellite altimeters than later years,
gaps with no eddy detected are present in both earlier and later
years.  An obvious observation from the inspection of the figure
is that trajectory lengths increase with $T$
increasing from 90 to 180 to 360 d.  Specifically,
these increase on average roughly from 450 to 900 to 1800 km.  This
is accompanied by a reduction in eddy size.  In effect, mean eddy
diameters decrease on average approximately from 140 to 100 to 50
km.  This suggests an average eddy decay rate of
40 km$^2$ d$^{-1}$ in 360 d.  Mean eddy translational
speeds remain quite stable around 5 km d$^{-1}$ (about twice the
speed of long baroclinic Rossby waves) independent of $T$.

We now proceed to discussing transport estimates.  Let $\Sigma$ be
a fixed curve and assume that is traversed by coherent material
eddies, episodically and all in the same direction.  We call
\emph{coherent transport} the contribution by such eddies to the
flux across $\Sigma$ of the two-dimensional velocity supporting the
eddies.  For a single eddy, this function takes nonzero values over
the time interval on which the eddy crosses $\Sigma$.  To a good
approximation such a function is given by a boxcar with amplitude
equal to the area of the eddy divided by the length of the interval.
For multiple eddies, a coherent transport time series will be given
by a sequence of boxcars with different amplitudes resulting from
superimposing episodic individual eddy contributions.

\begin{figure} 
  \includegraphics[width=\textwidth]{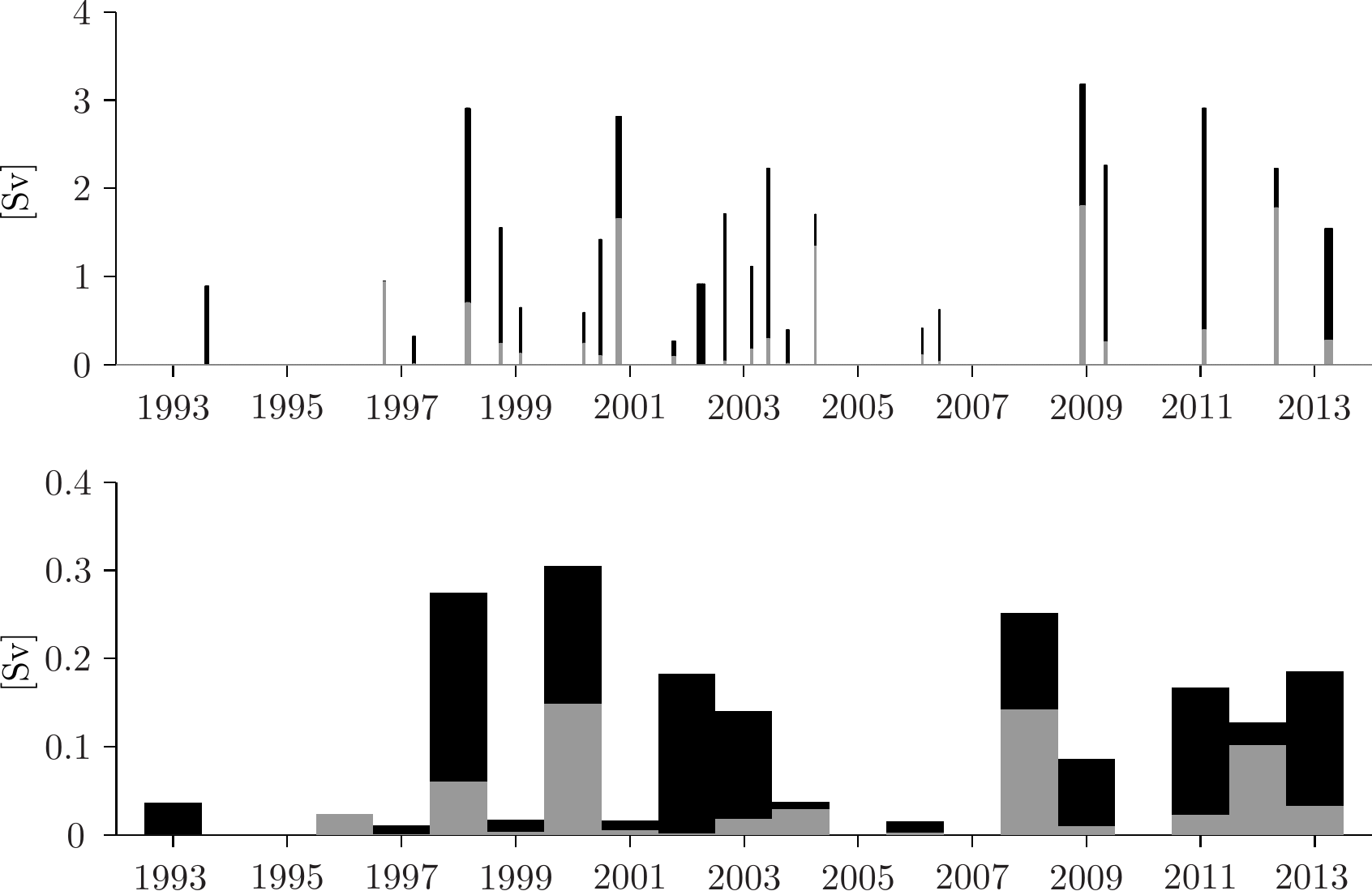}%
  \caption{Instantaneous (top panel) and annual average (bottom
  panel) time series of transport produced by 360-d coherent material
  eddies crossing the reference section indicated by the dashed
  segment in the bottom-left panel of \figref{fig:fig01}.  Gray-shaded
  bar portions correspond to transport of Indian Ocean water trapped
  inside the eddies.}
  \label{fig:fig02}%
\end{figure}

The top panel of \figref{fig:fig02} shows a time series of coherent
transport estimates over the period 1992--2013 obtained by considering
360-d eddies and $\Sigma$ as indicated by the dashed segment in
bottom-left panel of \figref{fig:fig01}, which is traversed by all
identified 360-d eddies (and their residuals, after coherence is
lost) in one direction.  These eddies have the ability to carry
water coherently for the longest distances, and thus are the most
meaningful for the transport computation.  While transport as defined
above is two-dimensional in nature, we report three-dimensional
transport values measured in Sv (1 Sv = $10^6$ m$^3$s$^{-1}$).
These were obtained by multiplying the computed two-dimensional
transport values by 1 km.  With a certain degree of
uncertainty \citep{deSteur-etal-04}, this value is in the range of
mature Agulhas ring trapping depths inferred from float-profiling
hydrography \citep{Souza-etal-11b}.  A distinguishing feature of
the computed transport time series is a large variability, both
intra- and interannual.  Nonzero transport estimates range approximately
from 0.25 to 3 Sv (about 1.5 Sv on average).  These varying transport
estimates are attributed mainly to varying eddy sizes.  Interspersed
zero transport gaps last roughly from 3 months to 3 years.  These
varying-length gaps cannot be explained by previously reported
Agulhas ring shedding rates of 1 ring every 2 to 3 months
\citep{Byrne-etal-95, Goni-etal-97, Schouten-etal-00, Souza-etal-11b}.
Rather, they are due to marked eddy episodicity.  Gray-shaded bar
portions in the top panel of \figref{fig:fig02} correspond to
transport of water that can be identified with leaking Indian Ocean
water into the South Atlantic.  The Indian Ocean water fraction
carried within eddies was estimated by advecting the eddy boundaries
in backward time for as long as at least 90\pct reversibility was
attained (about 1.25 years on average), and computing the proportion
of the enclosed fluid found east of 20$^{\circ}$E (the longitude
at which Indian Ocean and South Atlantic meet).  We note that
reversibility is strongly constrained by sensitive dependence on
initial conditions.  Also, the Agulhas retroflection typically
extends west of 20$^{\circ}$E.  Therefore, our estimates of Indian
Ocean water content should be considered as a lower
bound.  At least, then, the eddies are found this way to carry on
average about 30\pct of water that can be unambiguously identified
with Indian Ocean water.  Accordingly, the transport of Indian Ocean
water trapped inside the eddies is found to be  approximately 0.5
Sv on average.  Rare cases are eddies detected in mid 2002 and early
1996, which carry barely 1 and almost 99\pct of Indian Ocean water,
respectively.  The Indian Ocean water transported by these eddies
is about 0.01 and 1 Sv, respectively.

The bottom panel of \figref{fig:fig02} shows a time series of annual
coherent transport estimates computed by averaging the (instantaneous)
estimates in the top panel within each year (as in that panel,
gray-shaded bar portions correspond to Indian Ocean water transport).
The maximum annual transport produced by 360-d coherent material
eddies is  about 0.3 Sv.  Our estimate is two orders of magnitude
smaller than earlier estimates obtained as total volume of eddies
detected during a given year, divided by 1 year \citep{Garzoli-etal-99,
Richardson-07, Dencausse-etal-11, Souza-etal-11b}. This large
difference might be reduced by an order of magnitude if a larger
vertical extent for the eddies is assumed.  Indeed, \citet{vanAken-etal-03}
report vertical extents of 4 km, but only for very young Agulhas
rings.  The reason for this large discrepancy actually resides in
that these earlier estimates implicitly assume that eddies whose
diameters at detection time are 250 km or so can preserve material
coherence over periods as long as 1 year.  This cannot be guaranteed
by Eulerian analysis of altimetry or the inspection of in-situ and
profiling-float hydrography, and drifter and float trajectories,
which led to the earlier transport estimates.  Truly material eddies
as large as 250 km in diameter revealed from altimetry in the region
of interest can be guaranteed to preserve coherence for at most 3
months.  Beyond 3 months or so, eddies of this size shed filaments
that typically reach the generation region or further
east.  The
maximum annual transport of Indian Ocean water trapped inside 360-d
coherent material eddies does not exceed 0.2 Sv.  This is also
smaller, by two orders of magnitude, than annual Agulhas leakage
estimates obtained from numerical simulations \citep{Doglioli-etal-06,
Biastoch-etal-09a, LeBars-etal-14}.  While these Agulhas leakage
estimates still lack observational support, the noted large mismatch
suggests that eddies may not be as efficient in transporting leaked
Indian Ocean water across the South Atlantic as the early eddy
transport estimates appear to indicate.

\begin{figure}
  \includegraphics[width=\textwidth]{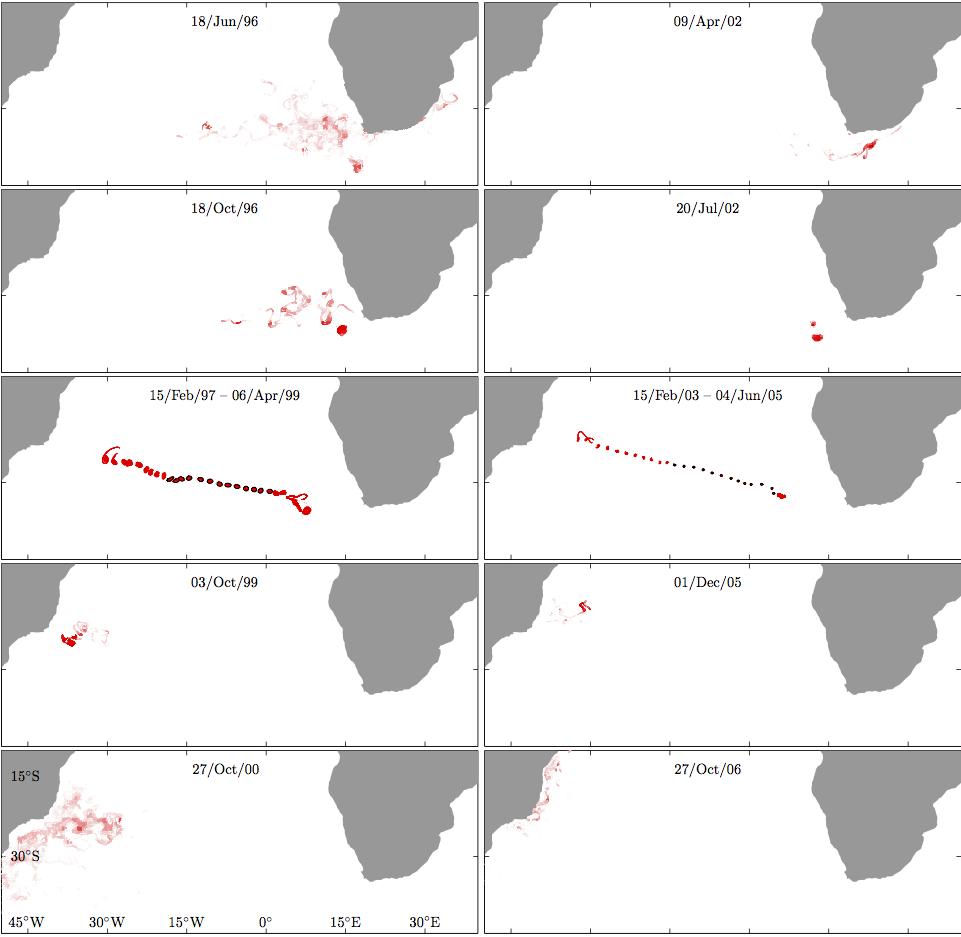}%
  \caption{Snapshots of the long-term evolution of two 360-d coherent
  material eddies with typical (left column) and exceptional (right
  column) genesis and demise stages.  The boundaries of the eddies
  while they constitute coherent material eddies are indicated in
  black.  Indicated in red are passive tracers that completely fill
  these eddies during their coherent material stage.}
  \label{fig:fig03}%
\end{figure}

Finally, we turn to discussing aspects of the evolution of detected
coherent material eddies. As before we focus on 360-d eddies, the
most persistent of all eddies detected.
Figure \ref{fig:fig03} shows snapshots of the long-term evolution
of two eddies selected according to their behavior during their
early and late evolution.  The eddy in the left column illustrates
typical behavior, while the eddy in the right column illustrates
exceptional behavior (animations including weekly snapshots are
supplied as Supporting Information files Movie S1 and Movie S2,
respectively).  The evolutions were constructed by advecting passive
tracers inside each eddy boundary (indicated in black) at detection
time in backward time (for as long as at least 90\pct reversibility
was attained) and also in forward time (beyond the theoretical
coherence time).  The typical behavior is characterized by organization
into small coherent material eddies (the specific eddy depicted in
the left column of \figref{fig:fig03} is
approximately 90 km in diameter) from rather incoherent fluid
composed of a mixture mainly of water that resides in the South
Atlantic and a much smaller fraction of water traceable into the
Indian Ocean.  Typical coherent material eddies emerge away from
the southern tip of Africa, just east of the Walvis Ridge in the
South Atlantic.  This genesis picture does not adhere to the commonly
accepted conceptual picture in which Agulhas rings are shed from
the Agulhas retroflection as a result of eventual Indian-Ocean-entrapping
occlusions \citep{Pichevin-etal-99}.  Our results are more consistent
with those from earlier works \citep{Schouten-etal-00, Boebel-etal-03}
reporting intense mixing in Cape Basin.  But coherence eventually
emerges from the mostly incoherent water resulting from this process
close to the  Walvis Ridge, and is followed by propagation of small
eddy cores with minor filamentation across the subtropical gyre.
Coherence is eventually lost, the contents of the eddies are mixed
with the ambient water in the vicinity of the bifurcation of the
subtropical gyre, and finally transported mostly southward by the
Brazil Current.  Out of a total of 23 eddies detected, 15 adhere to this picture.
The exceptional behavior, observed by only 1
eddy, is also characterized
by emergence of small coherent material eddies (the particular eddy
shown in the right column of \figref{fig:fig03} has roughly 40 km
in diameter) out of rather incoherent water, the only difference
being that most of the water inside this eddy is traceable into the
Indian Ocean.  This is still different from
the widely accepted ring genesis picture inasmuch as material
coherence is not acquired immediately but rather after some time,
near the Walvis Ridge.  Propagation across the subtropical gyre
then follows with minor filamentation until coherence starts to be
gradually lost.  The eddy contents mix with the surrounding water
near the bifurcation of the subtropical gyre.  The majority of these
are then transported northward close to the coast by the North
Brazil Current.  This behavior most closely adheres to the scenario
put forward by \citet{Gordon-Haxby-90}.  The behavior of the remaining
7 eddies detected share aspects
of the two markedly distinct behaviors just described.

\section{Conclusions}

Aided by a recently developed Lagrangian technique from nonlinear
dynamical systems theory, we have extracted from geostrophic
velocities derived from nearly two decades of altimetry measurements
a coherent transport signal across the South Atlantic through the
so-called Agulhas corridor.  The technique enables accurate,
frame-independent identification of mesoscale eddies with cores
whose material boundaries remain coherent, i.e., without showing
noticeable signs of filamentation, for up to one year.  These
coherent material eddies were used in the coherent transport
computation, which turned out to be
smaller (by at least two orders of magnitude) than earlier ring
transport estimates.  The main reason is that those transport
estimates implicitly assumed material coherence for eddies revealed
from their Eulerian footsteps in the altimetry dataset.  Such eddies
are either too large (by one order of magnitude) for long-range
material coherence to be guaranteed or simply do not represent
coherent material eddies.  The portion of Indian Ocean water
annually carried within the coherent material eddies
was also identified and found to be small (by two orders of magnitude)
compared to recent estimates of total Agulhas leakage based on
numerical simulations. This result suggests
a reduced role of Agulhas rings in transporting leaked Indian Ocean
water. We also investigated the evolution of the detected coherent
material eddies.  We found that the conceptual picture in which
Agulhas rings are shed from the Agulhas retroflection as a result
of episodic Indian-Ocean-water-entrapping occlusions is in general
not valid.  Coherent material eddies tend to emerge near the Walvis
Ridge from rather incoherent water, mostly residing in the South
Atlantic and to a small extent traceable into the Indian Ocean. How
this precisely happens is not known and is subject of ongoing
investigation.  The majority of the coherent material eddies formed
this way (15 out of a total of 23) were found to
cross the subtropical gyre and eventually to be
absorbed into the Brazil Current.  However, only 1
eddy consisting of mainly Indian Ocean water was
seen to pour its contents on the North
Brazil Current. This suggests that the contribution
by coherent material Agulhas rings to the global thermodynamical
budget is much less significant than originally thought
\citep{Gordon-Haxby-90}.  While we find ability of eddies to transport
water over long distances, this is less important than recently
argued for eddies beyond Agulhas rings based on lax assumptions
about material coherence \citep{Dong-etal-14, Zhang-etal-14}.
Finally, explicit resolution of three-dimensional aspects of ring
transport, which can be done by applying extensions of the technique
employed here \citep{Blazevski-Haller-14, Haller-14} to an ocean
general circulation model, are not expected to substantively modify
our results.  Indeed, these may be actually more constrained by
diffusion induced by submesoscale turbulence unresolved by altimetry
and most models.

\begin{acknowledgments}
  The constructive criticism of R.\ Abernathey and two anonymous
  reviewers led to improvements in the paper.  The altimeter products
  were produced by SSALTO/DUCAS and distributed by AVISO with support
  from CNES (http://\allowbreak www.\allowbreak aviso.\allowbreak
  oceanobs). Our work was supported by NSF grant CMG0825547, NASA
  grants NNX10AE99G and NNX14AI85G, and NOAA through the Climate
  Observations and Monitoring Program.
\end{acknowledgments}

\bibliographystyle{agu08}

\end{article}

\newpage

\vfill

\end{document}